\documentclass[fleqn,usenatbib]{mnras}

\usepackage{newtxtext,newtxmath}
\usepackage{txfonts}

\usepackage[T1]{fontenc}
\usepackage{ae,aecompl}

\usepackage{graphicx}	
\usepackage{amsmath}	
\usepackage{amssymb}	

\usepackage{xcolor}
\usepackage{url}						

\definecolor{crimson}{HTML}{DC143C}
\definecolor{darkred}{HTML}{8B0000}

\title[The frequency and central depletion of UDGs]{Reviewing the frequency and central depletion of Ultra-Diffuse Galaxies in galaxy clusters from the \texttt{KIWICS} survey\thanks{Based on observations made with the Isaac Newton Telescope operated on the island of La Palma by the Isaac Newton Group of Telescopes in the Spanish Observatorio del Roque de los Muchachos of the Instituto de Astrof\'isica de Canarias.}}

\author[Pavel E. Mancera Pi\~na et al.]{Pavel E. Mancera Pi\~na$^{1}$\thanks{e-mail: pavel@astro.rug.nl},
Reynier F. Peletier$^{1}$, 
J.A.L. Aguerri$^{2}$,
Aku Venhola${^{1,3}}$,\newauthor
Scott Trager$^{1}$ 
and Nelvy Choque Challapa$^{1}$ \\
\\
$^{1}$Kapteyn Astronomical Institute, University of Groningen, Lanleven 12, 9747 AD, Groningen, The Netherlands\\
$^{2}$Instituto de Astrof\'isica de Canarias, Calle V\'ia L\'actea S/N, La Laguna, Tenerife, Spain\\
$^{3}$Astronomy Research Unit, University of Oulu, FI-90014, Oulu, Finland
}

\date{Accepted XXX. Received YYY; in original form ZZZ}

\pubyear{2018}
\begin{document}
\label{firstpage}
\pagerange{\pageref{firstpage}--\pageref{lastpage}}
\maketitle

\begin{abstract}
The number of Ultra-Diffuse Galaxies (UDGs) in clusters is of significant importance to constrain models of their formation and evolution. Furthermore, their distribution inside clusters may tell us something about their interactions with their environments.
In this work we revisit the abundance of UDGs in a more consistent way than in previous studies. We add new data of UDGs in eight clusters from the Kapteyn IAC WEAVE INT Clusters Survey (\texttt{KIWICS}), covering a mass range in which only a few clusters have been studied before, and complement these with a compilation of works in the literature to homogeneously study the relation between the number of UDGs and the mass of their host cluster. Overall, we find that the slope of the number of UDGs--cluster mass relation is consistent with being sublinear when considering galaxy groups or linear if they are excluded, but we argue that most likely the behavior is sublinear.
When systematically studying the relation between the projected distance to the innermost UDG and M$_{200}$ for each cluster, we find hints that favor a picture in which massive clusters destroy UDGs in their centres.

\end{abstract}



\section{Introduction}

Ultra-Diffuse Galaxies (UDGs, \citealt{vandokkum}) are a extreme class of low surface brightness galaxies (LSB, e.g. \citealt{sandage}; \citealt{impey}; \citealt{conselice}) with dwarf-like surface brightness ($\mu_g$ \textbf{$\gtrsim$} 24 mag arcsec$^{-2}$) but L$^\star$-like effective radius ($R_e \gtrsim$ 1.5 kpc). They have colors of passively evolving stellar populations (although some of them, especially in the field, can host ongoing star formation), exponential-like S\'ersic profiles, and an axis ratio distribution with a peak around $b/a \sim$ 0.7$-$0.8 (e.g. \citealt{koda}; \citealt{vanderburg}; \citealt{roman}; \citealt{Aku}, \citealt{largepaper}).

In recent years UDGs have drawn a lot of attention because of their potential to test galaxy formation and evolution models at such extreme conditions. At the same time, the discovery of UDGs in a range of environments (e.g. \citealt{vandokkum}; \citealt{merritt}; \citealt{martinez}; \citealt{vanderburg}; \citealt{roman2}) represents a major opportunity to study the effects of environment on shaping galaxies.  

One of the first noticed characteristics of UDGs in galaxy clusters was the relation between the number of UDGs and the mass of their host cluster: the N(UDGs)--M$_{200}$\footnote{Here M$_{200}$ is used as a proxy of the cluster mass. It is defined as the mass enclosed by R$_{200}$, the radius at which the mean density is 200
times the critical density of the Universe.} relation (\citealt{vanderburg}, hereafter vdB+16). This relation is potentially very interesting to study the role of the environment affecting a UDG, since it gives information about the environment in which UDGs are preferentially found.
vdB+16 noticed a very tight relation: N(UDGs) $\propto$ M$_{200} ^{0.93 \pm 0.16}$, where N(UDGs) is the number of UDGs inside R$_{200}$. \citealt{roman2} (hereafter RT17b) extended this relation to galaxy groups and found N(UDGs) $\propto$ M$_{200} ^{0.85 \pm 0.05}$, a 3$\sigma$ sublinear relation. This slope implies that UDGs are more abundant, per unit host cluster mass, in low-mass systems. RT17b suggested that a slope less than one is an indication that UDGs either preferably form in low-mass groups, or they are more efficiently destroyed in very massive clusters, and it supports a picture of UDGs accreted from groups and/or the field to clusters, where some UDGs get destroyed due to interactions with the environment.
However, \cite{vanderburg2}, also studying the low-mass regime of the relation, found a slope of 1.11 $\pm$ 0.07, concluding that UDGs are more abundant, per unit cluster mass, towards more massive clusters.
The nature of this relation is thus not fully determined, and given its importance for our general understanding of UDGs and UDGs formation models (e.g. \citealt{amorisco}; \citealt{amoriscohaloes}; \citealt{carleton}), it is essential to reconcile these discrepancies, particularly whether the slope is linear or not. 

Another clue to understand the impact of the cluster environment on UDGs is their deficit in the inner regions of clusters (e.g. \citealt{vandokkum}; \citealt{merritt}; vdB+16; \citealt{wittmann}; \citealt{Aku}). 
While this could be a bias due to lower detectability in cluster centres, it is also possible that UDGs are unable to survive due to the strong potential forces (see for instance \citealt{merritt}, and the detailed analysis by \citealt{wittmann}). 
However, a consistent investigation of this effect with homogeneous data has not yet been undertaken.\\

\noindent
With the aim of understanding more about the formation and evolution of UDGs in galaxy clusters, we present here our results of a homogeneous analysis on both phenomena.
Using data of new UDG detections in eight galaxy clusters from \cite{largepaper}, hereafter Paper II, we perform a detailed comparison of our sample with UDGs in clusters in the literature. The rest of this work is organized as follows. In Section \ref{sec:sample} we present our data. In Section \ref{sec:results} we describe our findings regarding the abundance of UDGs and their central depletion, respectively. Finally, in Section \ref{sec:discussion} we discuss our results and summarise our conclusions. 

Along this work we use magnitudes in the AB system and we adopt a $\Lambda$CDM cosmology with $\Omega_\textnormal{m}$ = 0.3, $\Omega_{\Lambda}$ = 0.7 and H$_\textnormal{0}$ = 70 km s$^{-1}$ Mpc$^{-1}$.


\section{Cluster sample}
\label{sec:sample}
Our observations come from a deep photometric survey (PIs Peletier \& Aguerri) our team is carrying out of a set of X-ray selected, nearby (0.02 < $z$ < 0.04) galaxy clusters, which will be followed-up with the new WEAVE spectrograph \citep{dalton}: the Kapteyn IAC WEAVE INT Clusters Survey (\texttt{KIWICS}). For these observations we use the 2.5-m Isaac Newton Telescope of the Roque de los Muchachos Observatory on La Palma, Spain. The observations from \texttt{KIWICS} are ideal for studying the evolution of LSBs at low redshift, covering at least 1 $R_{200}$ (in projection) in each cluster, but the field of views are usually larger.

The images consist of deep $r$- (total integration time $\sim$1.5h) and $g$-band (total integration time $\sim$0.5h) observations, reduced using the \texttt{Astro-WISE} environment \citep{astrowise}. For illustration, the mean depth of the $r$-band in our whole sample is $\sim$29.3 mag arcsec$^{-2}$ when measured at a 3$\sigma$ level and averaged over boxes of 10 arcsec, comparable to the depth of many observations of UDGs in the literature (see RT17b). 
A detailed description of the observational strategy, data reduction processes, and the search of UDGs is given in Paper II, so we just briefly summarise the main aspects.

The sample consists of a set of eight, relatively well virialized and isolated clusters (see Table \ref{tab:sample}). We follow the strategy of detecting the potential UDG candidates using \texttt{SExtractor} \citep{sextractor} (based on their size and surface brightness and then fitting the galaxies with \texttt{GALFIT} \citep{galfit}, using the pipeline described in \cite{Aku} and \cite{Aku2} to retrieve the structural parameters. Simulations and sanity checks are done to determine the detection limits and completeness level of the sample, as to ensure its purity. {These simulations (cf. Figure 3 in Paper II) show that the completeness level for our sample is similar to vdB+16, and they help us to find an efficient way to run \texttt{SExtractor}, lowering the rate of false positives.
In Paper II we find 442 UDG candidates in these eight clusters, 247 being at projected clustercentric distances within R$_{200}$. The definition of UDG\footnote{We realize that by allowing high S\'ersic ($n < 4$) objects to be included we are allowing relatively concentrated objects, but, in agreement with the literature, we do not want to restrict our sample by excluding these objects a priori. In any case, the contribution of galaxies with $n \geq 2$ is < 3\%. The cut in color aims also to reject background objects that might look like UDGs but do not have colors representative of stellar populations of low-$z$ galaxies.} used in Paper II is galaxies with mean effective surface brightness $\langle\mu(r,R_e)\rangle \geq$ 24.0 mag arcsec$^{-2}$, effective radius $R_e \geq$ 1.5 kpc, S\'ersic index $n < 4$ and color $g-r < 1.2$ mag. All the galaxies are corrected for Galactic extinction (taken from \citealt{schlafly}), and while the effect is marginal given the redshifts of our sample, $k$-corrections (from \citealt{chilingarian}) and surface brightness dimming (from \citealt{tolman1,tolman2}) corrections are also taken into account. 
For the data description, the results about the structural parameters and scaling relations of UDGs, and their implications in understanding the formation and evolution of UDGs, please refer to Paper II. As a matter of illustration, Figure \ref{fig:examples} shows examples of some of the UDG candidates found in Paper II.

	\begin{table*}
	\caption{ID, coordinates, redshift, M$_{200}$, R$_{200}$, and number of UDGs with $R_{e,c} \geq$ 1.5 kpc for the clusters in our sample.}
	\label{tab:sample}
	\begin{center}
	\begin{tabular}{lccccccc}
	\hline
    Cluster & RA & DEC & Redshift & M$_{200}$ & R$_{200}$ & N(UDGs) & N(UDGs)\\
      & (hh:mm:ss) & ($^{\textnormal{o}}~:~^{'}~:~^{''}$) & &  ($\times$10$^{13}$ M$_\odot$)& (kpc) & raw & decontaminated\\ \hline
    \noalign{\smallskip}
     RXCJ1204.4+0154 & 12:04:25.2 & +01:54:02 & 0.0200 & 2.9 $\pm$ 0.9 & 630 $\pm$ 60   & 15  & 14 \\ \noalign{\smallskip}
        Abell 779    &  09:19:49.2 & +33:45:37 & 0.0231& 4.0 $\pm$ 1.2 & 700 $\pm$ 70 & 21  & 20\\ \noalign{\smallskip}
     RXCJ1223.1+1037 & 12:23:06.5 & +10:27:26 & 0.0256& 2.0 $\pm$ 0.6 & 550 $\pm$ 60    & 11  & 11 \\ \noalign{\smallskip}  
        MKW 4s & 12:06:37.4 & +28:11:01 & 0.0274& 2.3 $\pm$ 0.7 & 580 $\pm$ 60    & 5    & 5\\ \noalign{\smallskip}
     RXCJ1714.3+4341 & 17:14:18.6 & +43:41:23 & 0.0275& 0.6 $\pm$ 0.2 & 370 $\pm$ 40    & 7    &7\\ \noalign{\smallskip}
		Abell 2634   & 23:38:25.7 & +27:00:45 & 0.0312& 26.6 $\pm$ 8.0 & 1310 $\pm$ 130 & 60  & 55\\ \noalign{\smallskip}
        Abell 1177   & 11:09:43.1 & +21:45:43 & 0.0319 & 3.8 $\pm$ 1.1 & 690 $\pm$ 70 &  9  &8\\ \noalign{\smallskip}
        Abell 1314   & 11:34:50.5 & +49:03:28 & 0.0327& 7.6 $\pm$ 2.3 & 870 $\pm$  90 & 19   &16\\ 
      \hline
	\end{tabular}
	\end{center}
	\end{table*}
    
\begin{figure*}   
\centering
\includegraphics[scale=0.74]{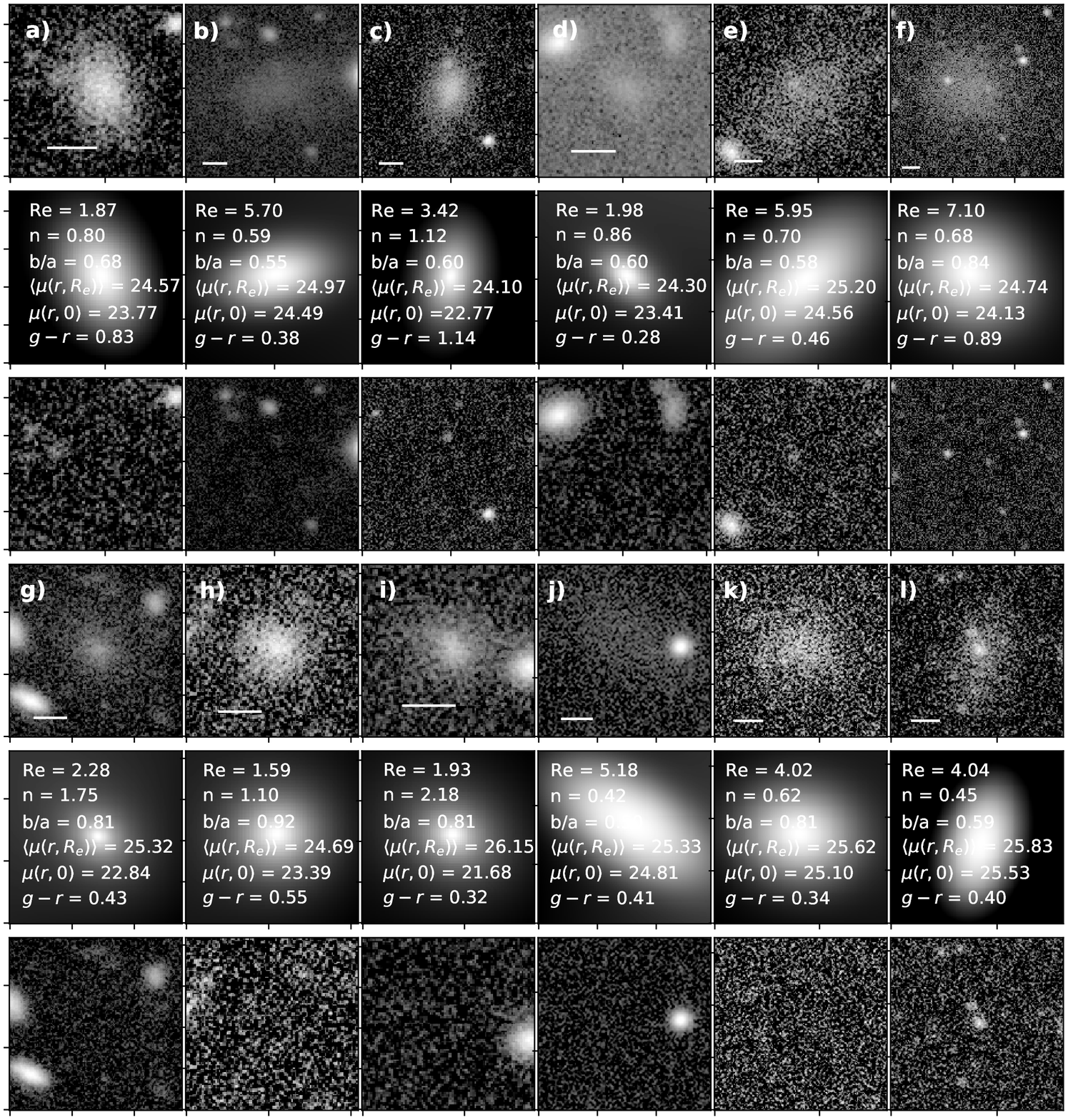}
\caption{Example of some UDG candidates found in Paper II. Top panels show the UDGs, mid panels the \texttt{GALFIT} models with the recovered structural parameters for each, and bottom panels the residuals. The white bars in the top boxes show a scale of 5 arcsec. The effective radii in mid panels are in kpc, surface brightness in mag arcsec$^{-2}$ and colors in magnitudes.}
\label{fig:examples}
\end{figure*}

\section{Results}
\label{sec:results}
\subsection{The frequency of UDGs in nearby clusters}
\label{sec:abundance}
In this section we aim to study the frequency of UDGs in clusters in a homogeneous way. We use our own dataset and complement it with literature data in a consistent way. We show how different considerations lead to different slopes for the relation, but overall, our analysis favors a sublinear behavior when galaxy groups are considered.

A careful and homogeneous analysis of the abundance of UDGs in the full explored range in cluster mass is still missing in most of the literature: when populating the N(UDGs)--M$_{200}$ plane, the numbers of UDGs given in each work are used directly, without taking into account the fact that the definition of a UDG is slightly different in the different papers.
Furthermore, there are several methods for determining the cluster mass, M$_{200}$, and this may have an impact in the relation.

We start by studying the relation only for the clusters of Paper II. We take the number of UDGs inside the projected R$_{200}$ of each cluster, and we statistically decontaminate it. The decontamination is done by analyzing observations of a blank field which was observed under the same conditions and strategy as the cluster sample and} following the same procedure (using \texttt{SExtractor} and \texttt{GALFIT} to select and characterize the UDG candidates) as for the cluster images, and we measure how many blank-field galaxies would have been classified as UDGs. 
The decontaminated number of UDGs in the cluster is then found by subtracting the expected contribution of interlopers from the number of UDGs found in the cluster. 
Since we will later compare with the literature, we consider UDGs with $R_{e,c}\geq$ 1.5 kpc\footnote{$R_{e,c}$ = $R_e \sqrt{b/a}$; this is slightly more restrictive than using the non-circularized effective radius, $R_e$, as in Paper II.}. 
It is worth mentioning that we analyzed with \texttt{GALFIT} blank-field galaxies larger than the angular size that a galaxy with effective radius of 1.5 kpc would have at $z = 0.7$, and therefore we can use the same blank-field galaxies to do background decontamination (for a similar data) of UDGs up to that redshift. Additionally, a second blank-field was observed and analyzed to check that the results are independent on which blank-field is used.

For each cluster, M$_{200}$ is derived in Paper II by fitting a Gaussian to the redshift distribution of the galaxies in the cluster (from the Sloan Digital Sky Survey (SDSS) and NASA/IPAC Extragalactic Database (NED) databases), estimating its corresponding velocity dispersion $\sigma$, and then using the $\sigma$--$M_{200}$ relation of \cite{munari}. 

\begin{figure}
\centering
\includegraphics[scale=0.627]{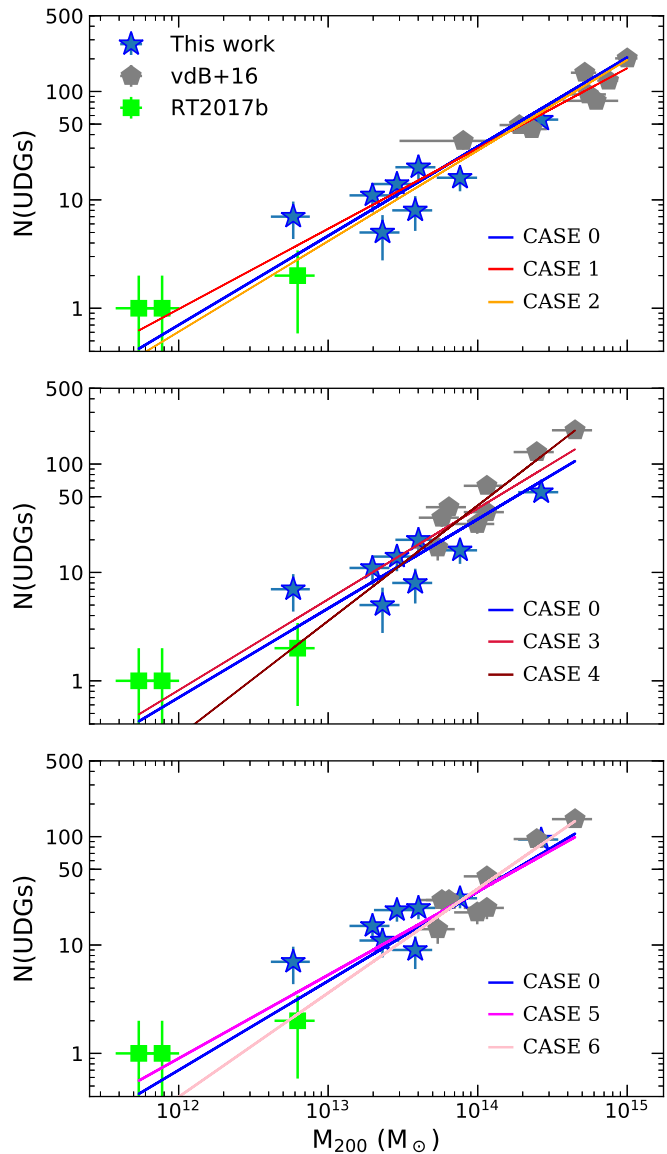}
\caption{Abundance of UDGs. \textit{Top:} The N(UDGs)--M$_{200}$ relation using the original M$_{200}$ values of the clusters in vdB+16. The corresponding fits, considering the 10$^{12}$ M$_\odot$ groups by RT17B (\textsc{case 1}) and without considering them (\textsc{case 2}), are shown with solid lines. \textit{Middle:} Same as left panel but considering our own mass determinations for the clusters in vdB+16 and the corresponding fits considering (\textsc{case 3}) and ignoring (\textsc{case 4}) the 10$^{12}$ M$_\odot$ groups. \textit{Bottom:} Abundance of UDGs considering only galaxies with $n < 2$, taking into account the 10$^{12}$ M$_\odot$ groups (\textsc{case 5}) and not taking them into account (\textsc{case 6}). See the text for details.}
\label{fig:abundance}
\end{figure}

Fitting\footnote{We use a Orthogonal Distance Regression fit, taking into account the uncertainties in both axis. The uncertainties in the y-axis are Poissonian, and come from considering uncertainties in the measurement and in the background subtraction.} the abundance relation for these eight clusters, shown in Figure \ref{fig:abundance}, we find N(UDGs) $\propto$ M$_{200} ^{0.82 \pm 0.24}$, a sublinear slope, although only at the $\sim$1$\sigma$ level\footnote{We note that the scatter in the relation by Paper II (i.e. considering non-circularized effective radii) is smaller, with a slope of 0.81 $\pm$ 0.17.}. This slope is consistent, within the uncertainties, with the slope by vdB+16 and RT17b. We call this \textsc{case 0}.

To expand the mass range of our study, we complement our sample with the samples of vdB+16 and RT17b. These two samples allow us to perform a homogeneous analysis these samples have similar depths and completeness, the methodology for the detection and characterization of UDG candidates was the same, and the luminosity and surface brightness distributions also resemble each other), as explained below.

First, vdB+16 selected UDGs with the same criteria as Paper II in surface brightness, but using R$_{e,c}$ > 1.5 kpc. Therefore their selection criteria is equivalent to ours. Their original MegaCam magnitudes are converted to our \texttt{SDSS} filters\footnote{We use the equations $g_{\textsc{mega}} = g_{\textsc{sdss}}-0.153\times(g_{\textsc{sdss}}-r_{\textsc{sdss}})$ and $r_{\textsc{mega}} = r_{\textsc{sdss}}-0.024\times(g_{\textsc{sdss}}-r_{\textsc{sdss}})$, as given in \url{http://www1.cadc-ccda.hia-iha.nrc-cnrc.gc.ca/community/CFHTLS-SG/docs/extra/filters.html}}, and we apply $k$- and surface brightness dimming corrections in the same way as for our sample. Finally, we keep galaxies with $b/a > 0.1$ and $-1 < g-r < 1.2$; this cut removes $\sim$ 5\% of the original sample. To decontaminate this sample, since the depth of the data is similar to ours, and the farthest cluster lies at $z$ < 0.07, we use the same blank field we used for our data, following exactly the same procedure. As a first guess, we use the M$_{200}$ and R$_{200}$ as reported by the authors in their paper.

Second, RT17b selected all the galaxies in their groups with $\mu(g,0) \geq$ 23.5 mag arcsec$^{-2}$ and $R_e$ > 1.3 kpc. Assuming a color $g-r$ = 0.6 and a S\'ersic profile $n=1$ (the mean color and S\'ersic index for UDGs according to Paper II), this corresponds to $\langle\mu(r,R_e)\rangle \geq$ 24.03 mag arcsec$^{-2}$. Therefore we assume that this sample is also complete for our analysis. We then take the parameters from the S\'ersic fit, and correct them for $k$-corrections and surface brightness dimming. 
These authors performed a very careful analysis looking for possible interlopers, and did not find any other LSB near their fields. Furthermore, they have two colors and their galaxies have both colors in agreement with spectroscopic members. Moreover, the groups are nearby ($z$ = 0.0141--0.0266) Hickson Compact Groups (HCGs, \citealt{hickson}) that by definition are isolated structures, and the galaxies are at relatively small projected distances from the centres of the groups. Finally, the association of several blue galaxies in RT17b with their corresponding HCG has been confirmed by spectroscopic observations \citep{spekkens}. For these reasons, we do not to apply extra background decontamination to this dataset. For M$_{200}$ and R$_{200}$, as in RT17b, we take the mean $\sigma$ values of the group and group+environment from \cite{hcgs}, and treat the data in the same way as ours. Of the eleven galaxies studied in RT17b, four fulfill our UDG definition and are at projected clustercentric distances < 1 R$_{200}$, so we include them in our analysis.

Two other papers would be particularly interesting to compare with: the groups by \cite{vanderburg2} and the very massive clusters by \cite{lee}. However, the characteristics and methodologies applied in those works are not fully consistent with the rest of data used here. In the case of \cite{vanderburg2}, i) their dataset is shallower by $\sim$ 0.5 mag than ours, ii) has no color constraints (which can increase the presence of interlopers; perhaps that also explains the relatively high S\'ersic indices they found), and iii) goes up to $z \sim$ 0.1, so the purity can be affected, the cosmological dimming is as high as $\sim$ 0.4 mag arcsec$^{-2}$, and the effects of having a PSF of the size of UDGs at $z \sim$ 0.01 might also play a role; all this may affect in different degrees the results by \cite{vanderburg2} but in any case our analysis is not fully compatible with that work. In the case of \cite{lee}, their clusters are at redshifts higher than what we can decontaminate with our blank field, and the extrapolation from the observed number of UDGs to the reported number inside R$_{200}$ is very large. Given these concerns we decided to not include those works for the sake of homogeneity.

We therefore have a homogeneous set of 19 systems, as shown in Figure \ref{fig:abundance}. As indicated in Table \ref{tab:slopes}, we find the fit N(UDGs) $\propto$ M$_{200}^{0.74 \pm 0.04}$, again a sublinear slope (hereafter \textsc{case 1}). \cite{vanderburg2} claimed that the $\sim$10$^{12}$ M$_\odot$ groups of RT17b may be not fully representative if most $\sim$10$^{12}$ M$_\odot$ haloes do not host UDGs. While this is not yet clear, for the sake of completeness we also fit the relation without taking into account the two lowest mass groups (\textsc{case 2}); this increases the slope to N(UDGs) $\propto$ M$_{200}^{0.84 \pm 0.07}$, still shallower than 1.

We also check the effect that different mass determinations have on the relation. In particular, the masses in vdB+16 come from the dynamical study by \cite{sifon} and probably suffer from different systematic effects than the masses of our sample or the rest of literature, since the membership criteria and $\sigma-$M$_{200}$ calibrations are different. To study this we derive the M$_{200}$ and R$_{200}$ for vdB+16 clusters in the same way as for our sample. The differences in the inferred masses are significant, with a median (mean) factor of 3.2 (3.8) and standard deviation 2, where the M$_{200}$ values are always smaller than the original dynamical masses. Taking this into account, we decide to perform two more fits considering the newly derived M$_{200}$ values for the vdB+16 sample, which of course affects R$_{200}$ and therefore N(UDGs). We also realize that the redshift distributions near these massive clusters are not as normally distributed as for our sample, something that perhaps might be affecting the purity of the sample.

In any case, if we now consider the 19 systems with the new mass measurements (\textsc{case 3}), we find, as for \textsc{case 2}, N(UDGs) $\propto$ M$_{200}^{0.83 \pm 0.07}$ (although the relation has a different intercept). Finally, we consider the new mass measurements without considering the two lowest mass groups (\textsc{case 4}), and this significantly increases the slope (and error) to N(UDGs) $\propto$ M$_{200}^{1.06 \pm 0.12}$. 
\begin{table*}
\caption{Slope of the N(UDGs)-M$_{200}$ relation for the cases described in the text. The second column refers to whether or not the mass used for the clusters in vdB+16 was the original, the third column indicates if the two lowest mass groups from RT17b are used, the fourth column specifies if only galaxies with $n < 2$ were used, and the last column gives the slope of the relation for each case.}
\label{tab:slopes}
\begin{center}
\begin{tabular}{lcccc}
	\hline
    \textsc{case} & M$_{200}$ homogeneous? & RT17b $\sim$10$^{12}$ M$_\odot$ groups? & constraint $n<2$? & slope\\ \hline
    \noalign{\smallskip}
   \textcolor{black}{vdB+ 16} & yes & no & no & 0.93 $\pm$ 0.16\\ \noalign{\smallskip}
      \textcolor{black}{RT17b} & no & yes & no & 0.85 $\pm$ 0.05\\ \noalign{\smallskip}
   \textcolor{black}{vdB+ 17} & no & no & no & 1.11 $\pm$ 0.07\\ \noalign{\smallskip}
         \textcolor{blue}{\textsc{case 0}} & --- & --- & no & 0.82 $\pm$ 0.24\\ \noalign{\smallskip}
    \textcolor{red}{\textsc{case 1}} & no & yes & no &0.74 $\pm$ 0.04\\ \noalign{\smallskip}
    \textcolor{orange}{\textsc{case 2}} & no& no & no &0.84$\pm$ 0.07 \\ \noalign{\smallskip}
    \textcolor{crimson}{\textsc{case 3}} & yes & yes & no &0.84 $\pm$ 0.07\\ \noalign{\smallskip}
   \textcolor{darkred}{ \textsc{case 4}} & yes & no& no &1.06 $\pm$ 0.12\\ \noalign{\smallskip}
       \textcolor{magenta}{\textsc{case 5}} & yes & yes & yes &0.77 $\pm$ 0.06\\ \noalign{\smallskip}
   \textcolor{pink}{ \textsc{case 6}} & yes & no& yes &0.96 $\pm$ 0.11\\ 
    \hline
\end{tabular}
\end{center}
\end{table*}

Motivated by the almost non-existent population of highly resolved UDGs with S\'ersic index > 2 (e.g. RT17b; \citealt{trujillo}; \citealt{Aku}; \citealt{cohen}), it is worth exploring how the abundance relation behaves if we consider of our analysis only galaxies with $n < 2$. We study this in \textsc{case 5} and \textsc{case 6}, considering and not the 10$^{12}$ M$_\odot$ groups, respectively. The result is shown in the right hand panel of Figure \ref{fig:abundance}. As expected, since the sample of vdB+16 contains a higher contribution of galaxies with $n > 2$ than ours, the new constraint lowers the slope of the relation. As stated in Table \ref{tab:slopes}, \textsc{case 5} has a slope of 0.77 $\pm$ 0.06, while \textsc{case 6} has a slope of 0.96 $\pm$ 0.11.


Overall, our analysis shows the importance of applying the same selection criteria when studying the abundance of UDGs, as well as in the mass estimations. It also indicates the dependence of the slope on the cluster mass regime considered, as we discuss in Section \ref{subsec:disc1}.

\subsection{The lack of UDGs in the centre of clusters}
\label{sec:depletion}
\begin{figure*}
\centering
\includegraphics[scale=0.68]{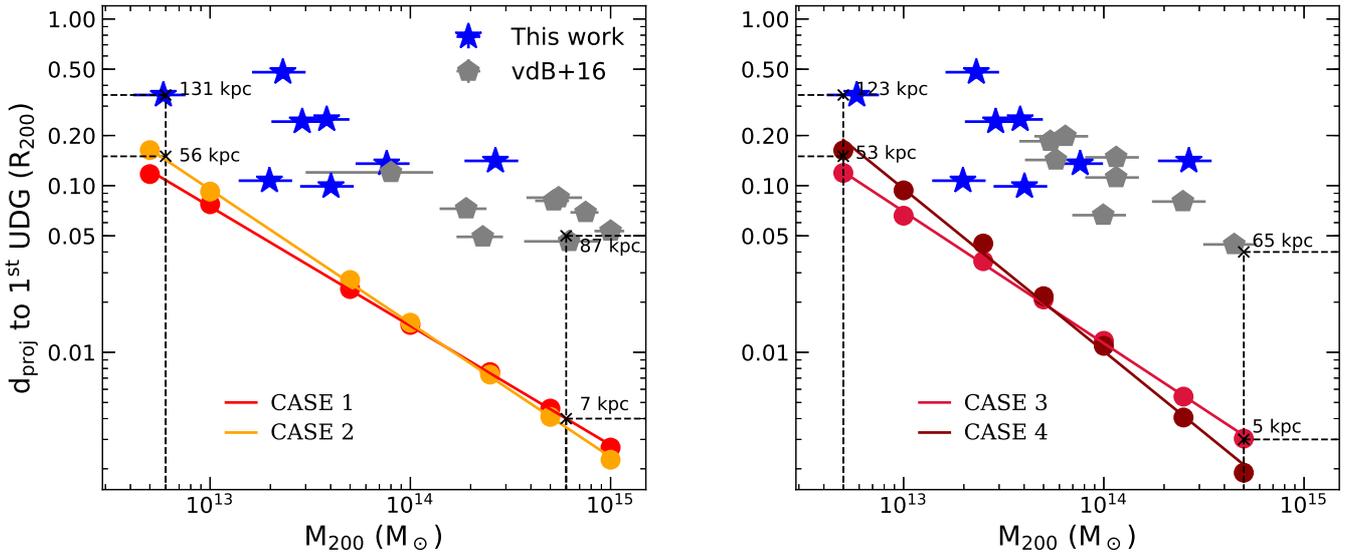}
\caption{Projected distances to the innermost UDGs as a function of the host cluster mass, M$_{200}$. Left panel shows the relation for our sample and the sample of vdB+16 with their original mass determinations, while the right panel shows the same but for our mass estimation of their clusters. The predicted positions using the Einasto profile and the different cases (different N(UDGs)-M$_{200}$ relations) are shown for each panel. The crosses and numbers in black show the physical distances corresponding to the distance and cluster mass indicated by the dotted lines, for clusters at $z=0$. See the text for details.}
\label{fig:dist2first}
\end{figure*}

To study the lack of UDGs in the innermost regions of clusters, we look at the projected distances at which the innermost UDGs appear. 
For this, we plot these distances as a function of the cluster mass in Figure \ref{fig:dist2first} (for the two mass estimations for the clusters from vdB+16). As can be seen, a striking relation appears, where UDGs in low-mass systems\footnote{The groups by RT17b are not used here, for the sake of consistency with the derived surface density profile; see below.} appear at larger distances (relative to R$_{200}$). While an initial conclusion that low-mass groups destroy the UDGs in their inner regions (supported also theoretically, cf. \citealt{mihos}) can be made, this apparent effect is an artefact: the probability of finding a UDG at any position is higher for more massive clusters, because they have more UDGs than groups.
Therefore, we decide to compare these empirical points with a simple prediction based on what we could expect from the observed distribution of UDGs. 

vdB+16 used an Einasto profile \citep{einasto} to characterize the radial surface density profile of UDGs, demonstrating that it produces a reasonable fit to their data.
In Paper II the surface density distribution of our sample is also studied, finding strong similarities between our profile and the profile from vdB+16.
Motivated by this similar behavior, we combine both profiles to build a general Einasto profile for the UDGs in both samples. The details about the derivation of this profile can be found in Paper II.

After deriving the general Einasto profile, we convert it to a probability function. Subsequently, we take a range of values in M$_{200}$ and inject the expected number of UDGs according with the N(UDGs)--M$_{200}$ relation (for the different cases mentioned above, excluding \textsc{case 0}, which only considers our data), at random positions that, however, follow the probability function of the surface density profile, extrapolated until the cluster centres. The result of this experiment in Figure \ref{fig:dist2first} shows that the trend remains, with the ratios between the observed and the Einasto-derived distances increasing towards the high-mass clusters. We discuss the possible implications of this result in Section \ref{subsec:disc2} below.



\section{Discussion and conclusions}
\label{sec:discussion}

\subsection{The frequency of UDGs}
\label{subsec:disc1}
As we have shown by using more clusters analyzed in a homogeneous way, different data used to infer the N(UDGs)--M$_{200}$ relation imply different slopes.
Given that we are sure about the high-purity of our sample, and for the sake of homogeneity, we consider in principle \textsc{cases 0, 3} and \textsc{4} as the most relevant. While \textsc{case 0} is consistent with the other two, \textsc{cases 3} and \textsc{4} are statistically different. This shows that the selection of the mass range determines the behavior of the relation: if one considers groups down to $\sim$10$^{12}$ M$\odot$ (RT17b) the slope is sublinear, but otherwise it is in agreement with being linear. 
If one takes into account only galaxies with S\'ersic index smaller than 2, as in \textsc{case 5} and \textsc{case 6}, then the slope becomes sublinear regardless the mass regime considered, although the uncertainties of \textsc{case 6} make it consistent with linear too. It is also worth mentioning that these slopes are all in agreement with the observed abundance of dwarfs in clusters: 0.91 $\pm$ 0.11 \citep{dwarfs}.
Notwithstanding, despite the results by \cite{vanderburg2} (whose limitations have been explained above), several studies of deep imaging in low-density environments (e.g. \citealt{merritt}; \citealt{martinez}; RT17b; \citealt{muller}; \citealt{cohen}; Paper II) have found the presence of LSBs and UDGs. We take this as evidence that the 10$^{12}$ M$\odot$ groups of RT17b are rather representative, and thus the slope of the abundance relation for UDGs is more likely to be sublinear, as in \textsc{case 3} or as in \textsc{case 5} if one imposes the extra constraint of a small S\'ersic index. Moreover, a selection bias present in most of the literature on UDGs should be taken into account: blue UDGs are brighter than red UDGs, which allows them to escape the surface brightness criterion used to define a UDG, as discussed in \cite{trujillo} (for instance, of the eleven galaxies studied in RT17b, only five meet our definition of UDG); this implies that blue analogues of the UDG population are systematically missed. Given that low-density environments have a larger contribution of blue galaxies than high-density environments, it is clear that the selection bias affects galaxy groups more strongly than massive galaxy clusters. Therefore, the slope of the N(UDGs)--M$_{200}$ relation as studied here and in the literature can be seen as an upper limit, and taking into account the contribution of bluer analogues the slope of the abundance relation would be even more sublinear.

As discussed in RT17b and \cite{vanderburg2}, a sublinear behavior implies that UDGs are more abundant, per unit host cluster mass, in low-mass systems. This could happen if UDGs preferably form/survive more easily in groups, or if they are destroyed in high-mass clusters. An alternative could be that the subhalo mass distribution of UDGs \citep{amoriscohaloes} is different for clusters of different mass, assuming the halo mass function is approximately universal \citep{jenkins}; then linearity would not be expected. 
A combination of all the above scenarios is of course possible, but with our current data we are not able to distinguish between them. 

\subsection{The depletion of UDGs in the centre of clusters}
\label{subsec:disc2}
A number of works have suggested that the absence of UDGs in the centre of clusters is due to UDGs not being able to survive the strong tidal forces (e.g. \citealt{vandokkum,merritt}; vdB+16). In particular, \cite{wittmann} discussed the topic in detail, arguing in favor of this scenario.
Moreover, as summarised by \cite{smith}, simulations often study the effects of the cluster potential as scaling with the projected distance to the inverse cubed, and with the size of the galaxy to the third power (e.g. \citealt{byrd}). This means that the innermost galaxies in clusters are expected to be more affected, and given that UDGs are large galaxies, are even more prone to this. Moreover, as those authors mention, harassment \citep{moore} can also cause tidal mass loss, and low surface brightness disk galaxies are more susceptible to this loss (see also \citealt{gnedin}).
Given their clear absence in basically all the clusters studied in the literature, and considering the typical sizes of the central bright cluster galaxies (BCGs) it is likely that the observed lack of UDGs is not only explained by an observational bias: even for the most massive clusters studied here, the expected half-light radius of a BCG is around $\sim$ 5-20 kpc (following \citealt{laporte} and \citealt{sizes}), but innermost UDGs appear at larger projected clustercentric distances.

In our systematic study of the lack of UDGs in the innermost regions of clusters, we find hints of the central depletion of UDGs being caused by their destruction in the most massive clusters: the differences in the predicted (from the random placement of UDGs in an Einasto distribution) and observed distances to the innermost UDGs deviate systematically towards high-mass systems. Our simulations are rather schematic since, for instance, we treat UDGs as point sources, but they give an idea of the expected positions if more physical processes were not involved. A more realistic approach would be injecting mock UDGs with their expected structural parameters, and that follow the observed radial surface density distribution, in a set of different cluster images with a diversity of BCGs, and look then for the innermost UDGs but such analysis is out of the scope of this paper.
As discussed, potential-driven forces and harassment are likely to be behind the origin of UDGs avoiding the cluster centres.  However, considering the predicted distances for the most massive clusters, galactic cannibalism may be also playing a role: for a 10$^{15}$ M$_\odot$ cluster, the expected distance is $\sim$ 5-10 kpc, which is the order of the size of the BCG in that kind of massive cluster. This could be one of the mechanisms making the slope of the abundance relation sublinear. Moreover, if the slope is rather linear, there should be a mechanism effective in massive clusters that is restoring the linearity by creating more UDGs.\\


\noindent
To summarize our results, using new observations of UDGs in eight clusters, and complementing them with literature data, we performed a homogeneous analysis to study the abundance of UDGs and its central depletion in galaxy clusters. Our analysis shows the sensitivity that the slope of the N(UDGs)--M$_{200}$ relation has on the data used to derive it. 
Based on the current evidence we support a sublinear behavior for the relation, but we show the effects that different constraints have on the result. We found hints of one mechanism that could be making the abundance slope sublinear: from looking at the projected distance to the innermost UDG, we noticed that the deficit of UDGs increases with the cluster mass, supporting the idea that environmental effects are destroying UDGs in the central regions of high-mass systems.



\section*{Acknowledgements}

We thank the constructive comments by an anonymous referee, that helped to improve this paper.
We thank Remco van der Burg for sharing the data of vdB+16 with us, as well as for many clarifications on it. We also thank Javier Rom\'an for the data of RT17b and for enlightening discussions about our results. PEMP thanks the Netherlands Research School for Astronomy (NOVA) for the funding via the NOVA MSc Fellowship. PEMP, RFP and AV acknowledge financial support from the European Union's Horizon 2020 research and innovation programme under Marie Sk\l odowska-Curie grant agreement No. 721463 to the SUNDIAL ITN network. JALA acknowledges support from the Spanish Ministerio de Econom\'ia y Competitividad (MINECO) by the grants AYA2013-43188-P and AYA2017-83204-P. AV would like to thank the Vilho, Yrj\"o, and Kalle V\"ais\"al\"a Foundation of the Finnish Academy of Science and Letters for the funding during the writing of this paper. We have made an extensive use of SIMBAD and ADS services, as well as of the Python packages NumPy \citep{numpy}, Matplotlib \citep{hunter} and Astropy \citep{astropy}, for which we are thankful.













\bsp	
\label{lastpage}
\end{document}